\documentclass[11pt,twoside,dvips]{article}
\usepackage{verbatim}
\usepackage{amssymb}
\usepackage{amsfonts}
\usepackage{amsmath}
\usepackage{pifont}
\usepackage[spanish,english]{babel}
\usepackage[pdftex]{graphicx}
\usepackage{verbatim}
\usepackage{epsfig}
\usepackage{bm}
\usepackage{longtable}
\usepackage{fancyhdr}
\usepackage{multirow}

\begin{document}

\fancypagestyle{plain}{%
\fancyhf{}%
\fancyhead[LO, RE]{XXXVIII International Symposium on Physics in Collision, \\ Bogot\'a, Colombia, 11-15 september 2018}}

\fancyhead{}%
\fancyhead[LO, RE]{XXXVIII International Symposium on Physics in Collision, \\ Bogot\'a, Colombia, 11-15 september 2018}

\title{Recent results from Long-Baseline Neutrino experiments}
\author{Mario A. Acero$\thanks{%
e-mail: marioacero@mail.uniatlantico.edu.co}$, (for the NOvA Collaboration) \\ Programa de F\'isica, Universidad del Atl\'antico, \\ Carrera 30 8-49 Puerto Colombia, Colombia}
\date{}
\maketitle

\begin{abstract}
Understanding the physics of neutrinos is of paramount relevance for the development of high energy physics, cosmology and astrophysics, thanks to their characteristics and phenomenology. In particular, the property of changing flavor while neutrinos travel, the so-called neutrino oscillation phenomenon, provides us with valuable information about their behavior and their impact on the standard model of particles and the evolution of the universe, for instance. 
\\
Here I present an overview of the most recent results as reported by relevant experiments studying neutrinos produced by accelerator facilities and detected after traveling long distances: the so-called Long-Baseline neutrino experiments.
\end{abstract}

\section{Introduction}
It is a very well stablished fact that neutrinos are massive particles (contrary to what the Standard Model --SM-- suggests) and that mix: neutrino-flavor eigenstates are related to neutrino-mass eigenstates though the mixing (also known as the PMNS) matrix, as
\begin{equation}\label{eq_nuMix}
\left(
\begin{array}{c}
\nu_e      \\
\nu_{\mu}  \\
\nu_{\tau} \\
\end{array}
\right) 
= U_{\rm{PMNS}}
\left(
\begin{array}{c}
\nu_1      \\
\nu_2  \\
\nu_3 \\
\end{array}
\right).
\end{equation}
Here, the vector on the left represents the three flavor neutrinos ($\nu_{\alpha}, \alpha=e,\mu,\tau$) which are actually created and detected (through weak interaction processes), while the vector on the right includes the definite-mass neutrinos ($\nu_i, i=1,2,3$), which propagate through vacuum or matter.

The mixing matrix in (\ref{eq_nuMix}) is usually parametrized in terms of three orthogonal rotation matrices, each one depending upon the three so-called mixing angles ($\theta_{ij}, i\neq j=1,2,3$), and a phase\footnote{Here, neutrinos are assumed to be Dirac particles.} which parametrizes the CP violation in the lepton sector, ($\delta_{CP}$):
\begin{equation}
\left(
\begin{array}{c}
\nu_e      \\
\nu_{\mu}  \\
\nu_{\tau} \\
\end{array}
\right) 
= R(\theta_{23}) \cdot R(\theta_{13},\delta_{CP}) \cdot R(\theta_{12})
\left(
\begin{array}{c}
\nu_1      \\
\nu_2  \\
\nu_3 \\
\end{array}
\right).
\end{equation}
The mixing angles and the CP-violating phase, together with two mass-squared difference ($\Delta m^2_{21}$, $\Delta m^2_{31}$; $\Delta m^2_{jk} \equiv m_j^2 - m_k^2$), define the change of flavor that neutrinos can undergo while traveling an specific distance (from the source to the detector, for instance), a phenomenon known as neutrino oscillations.

A number experiments have studied neutrino oscillations using different neutrino sources (the sun \cite{Ahmad:2002jz}, the atmosphere \cite{Fukuda:1998mi}, nuclear reactors \cite{Eguchi:2002dm,An:2012eh,Ahn:2012nd}, accelerators \cite{Abe:2011sj,Adamson:2017gxd}) and detection techniques, measuring five out of six of the parameters with great precision. However, there are still important open questions regarding this phenomenon, having implications in other areas of particle physics, cosmology and astrophysics: it is not yet clear wether there is maximal mixing in the $\mu-\tau$ sector or not (i.e., is $\theta_{23} = \pi/4$?); we do not know the value of $\delta_{CP}$ (i.e. is there a violation of CP symmetry in the lepton sector?); the neutrino mass pattern (order or hierarchy) is unknown (i.e. is $\Delta m^2_{32}>0$ --normal order-- or $\Delta m^2_{32}>0$ --inverted order--?). 

There are other important and interesting questions\footnote{Are there only three neutrinos? What is the absolute mass of the neutrinos? Is the neutrino its own antiparticle?}, but answer to those just exposed are currently under investigation by some of the long-baseline (LBL) neutrino experiments. In the following sections, a short review of the recent results from accelerator-based oscillation neutrino experiments is presented.
\section{LBL neutrino experiments}
Results from four LBL accelerator-based neutrino experiments are shown here: NOvA, T2K, MINOS and OPERA. All of them use a neutrino beam created by a similar mechanism: high energy protons are fired against a fixed target (made of graphite or beryllium, for instance), producing $\pi^+$ ($\pi^{-}$) which, after decaying, generate a beam mainly composed by muon (anti)neutrinos.

So produced muon-(anti)neutrinos, traveling across the earth, may change flavor with a probability which depends on the mixing angles and the mass-squared differences. In this way LBL experiments are able to study neutrino physics from the observation of $\nu_{\mu}$ disappearance, $\nu_{e}$ appearance, and $\nu_{\tau}$ appearance, allowing them to measure the oscillation parameters (mainly $\theta_{13}, \theta_{23}, \Delta m^2_{32}, \delta_{CP}$, and and to study possible differences between neutrinos and antineutrinos.
\subsection{NOvA}
The NuMI\footnote{Neutrinos at the Main Injector} Off-axis $\nu_e$ Appearance (NOvA) experiment \cite{Ayres:2007TDR,Adamson:2017gxd,NOvA:2018gge} is a two-detector accelerator-based neutrino experiment designed to study the appearance of electron-(anti)neutrinos from a beam of muon-(anti)neutrinos. The $\nu_{\mu}$ beam travels through the earth from the Near Detector (ND) (100 m underground) at Fermilab, to the 14 kton Far Detector (FD) in Ash River, Minnesota, around 810 km apart. The FD is located 14 mrad off the centerline of the neutrino beam coming from Fermilab, so that the flux of neutrinos has a narrow peak at an energy of 2 GeV, the energy at which oscillation from muon neutrinos to electron neutrinos is expected to be at a maximum.

After collecting data from neutrinos and antineutrinos beams, NOvA has observed 58 neutrino (15 background) and 18 (5 background) antineutrino events, while studying the neutrino $\nu_e$ appearance,
thus providing a larger than $4\sigma$ evidence of electron anti-neutrino appearance. From the $\nu_{\mu}$ disappearance analysis, NOvA observed 113 neutrino and 65 antineutrino events, when 730 and 266 events were expected in absence of oscillations, respectively \cite{Sanchez:2018Neutrino,Vahle:2018Nufact}.
\begin{figure}[!htb]
\begin{center}
  \includegraphics[scale=0.35,trim=0 55 0 0,clip]{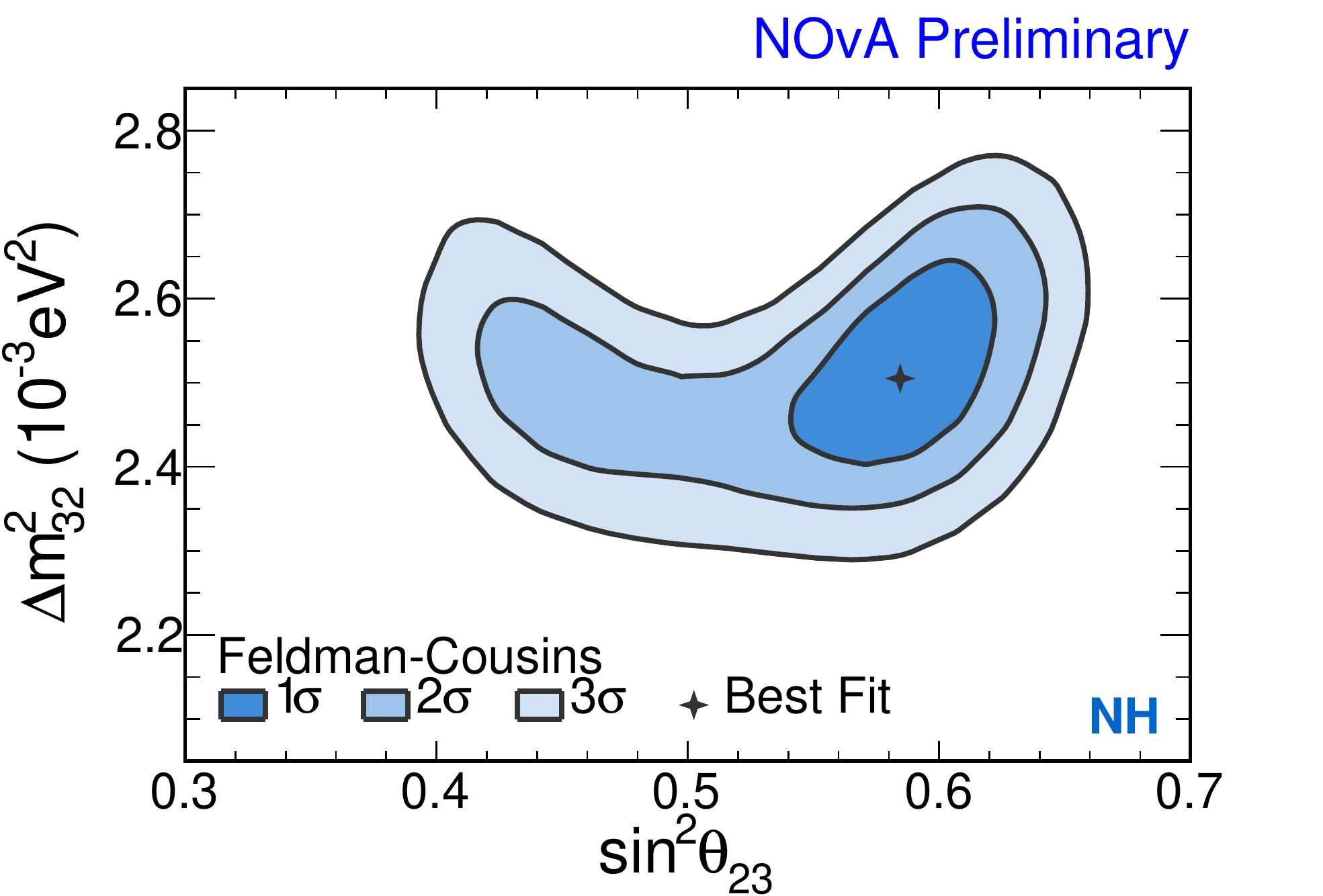}
  \includegraphics[scale=0.35,trim=0 55 0 0,clip]{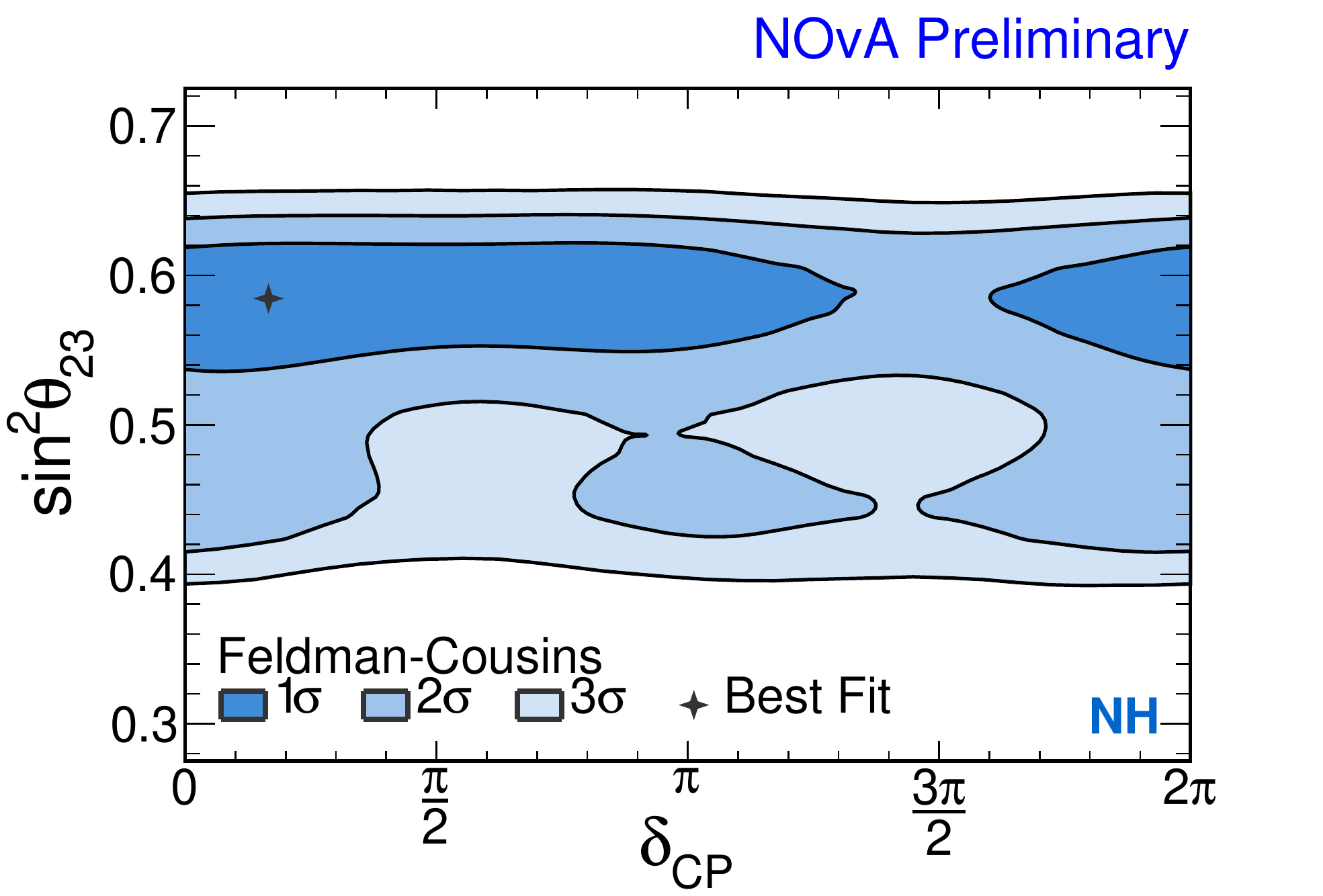}
  \includegraphics[scale=0.35,trim=0 0 0 30,clip]{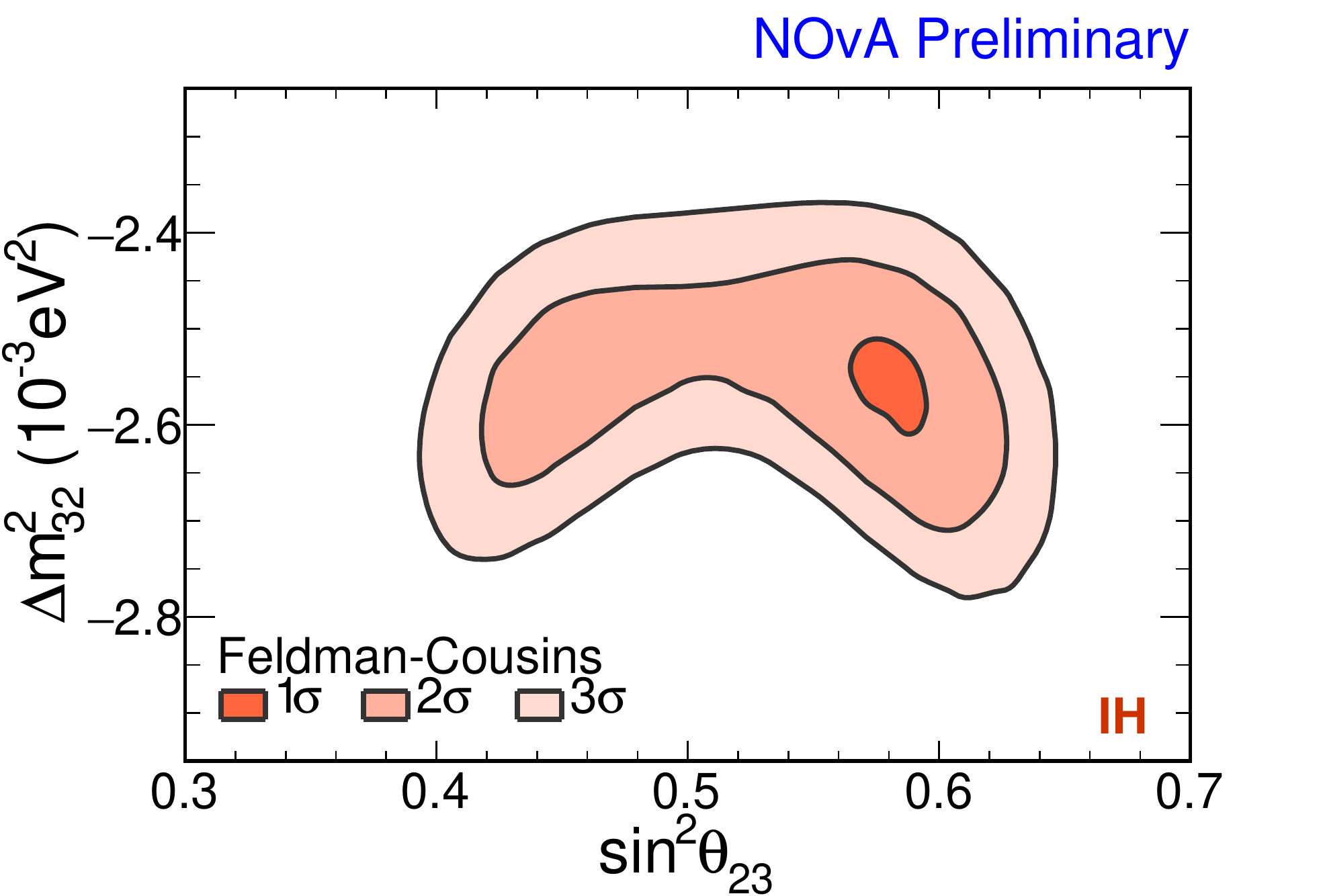}
  \includegraphics[scale=0.35,trim=0 0 0 30,clip]{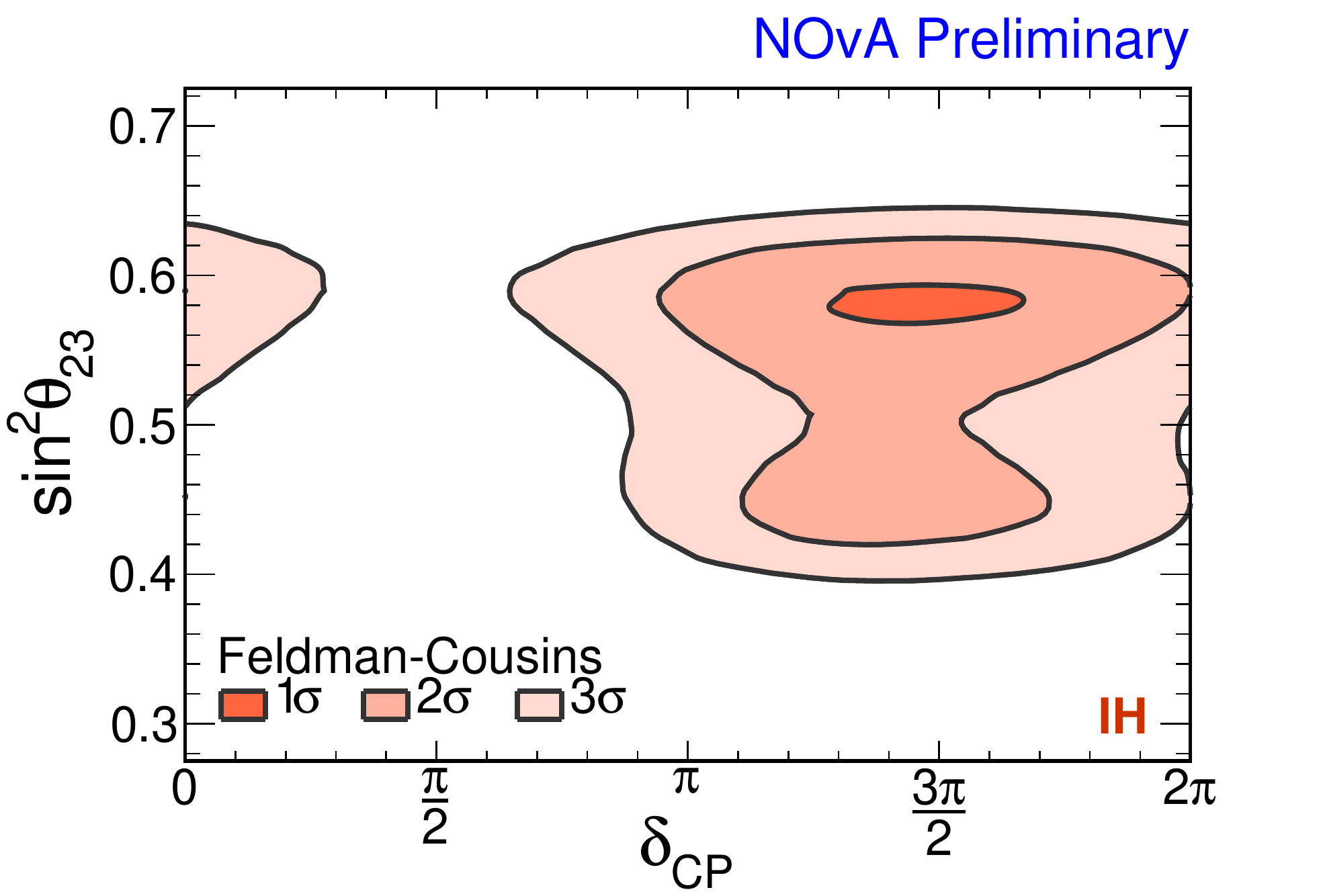}
  \caption{\label{fig_NOvA_contours}
    Regions of $\Delta m^2_{32}$ vs.~$\sin^2\theta_{23}$ (\emph{left}) and $\sin^2\theta_{23}$ vs.~$\delta_{CP}$ parameter spaces obtained from the $\nu_e$-appearance and $\nu_{\mu}$-disappearance data analysis at various levels of significance. The top panels correspond to normal mass ordering while the bottom panels to inverted ordering \cite{Sanchez:2018Neutrino,Vahle:2018Nufact}.}
\end{center}
\end{figure}

A joint $\nu_{\mu}$-disappearance--$\nu_e$-appearance data analysis allows NOvA to constraint the oscillation parameters as depicted in Fig.~\ref{fig_NOvA_contours}, where 1,2 and 3$\sigma$ C.L.~allowed regions for both, normal (top panels) and inverted (bottom panels) mass orderings are shown for comparison. The best fit values of the oscillation parameters are
\begin{equation}
\Delta m^2_{32}   = 2.51^{+0.13}_{-0.08} \times 10^{-3} \rm{ eV}^2, 
\qquad
\sin^2\theta_{23} = 0.58 \pm 0.03,
\qquad
\delta_{PC}       = 0.17\pi.
\end{equation}
In addition, NOvA data favor the normal mass ordering, non-maximal mixing with $\theta_{23} > 45^{\circ}$, and excludes $\delta_{CP} = \pi/2$ at more than $3\sigma$ C.L.~for the inverted mass ordering.
\subsection{T2K}
T2K\footnote{Tokai-to-Kamioka} is a LBL neutrino experiment \cite{Abe:2011ks} studying neutrino oscillations using muon (anti)neutrinos produced at the Japan Proton Accelerator Research Center (JPARC). The neutrino beam is directed towards a two detector system: one located 280 m from the production point, and the other, far detector, located 295 km away, at the Kamioka Observatory, 2.5$^{\circ}$ off-axis with respect to the neutrino beam \cite{Khabibullin:2018bmg} (neutrino energy spectra peaked at 800 MeV).

The T2K analysis was performed using data from neutrinos and antineutrinos and studying $\nu_{\mu}$-disappearance as well as $\nu_{e}$-appearance channels. After comparing the observed rates at the FD to predictions under oscillation hypothesis, they found that data is consistent with that model, for any value of $\delta_{CP}$. However, regarding the $\bar{\nu}_e$-appearance, T2K observed fewer events than expected for any value of $\delta_{CP}$ (9 events observed while 11.8 events were expected with oscillations and 6.5 without oscillations), preventing them to arrive to a robust statistical conclusion.
\begin{figure}[!htb]
\begin{center}
\includegraphics[scale=0.35,trim=0 0 0 0,clip]{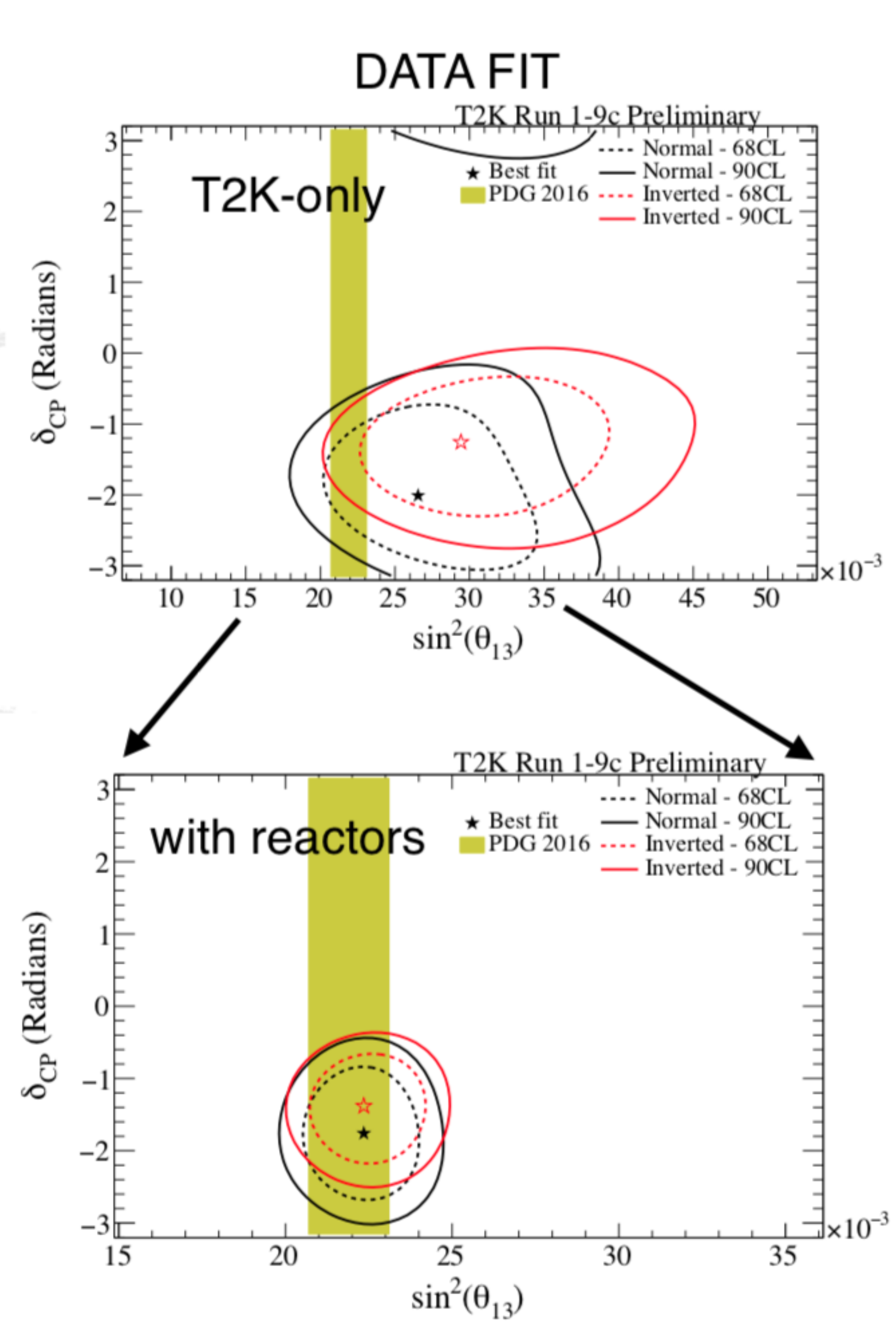}\vspace*{-9.5cm}
  \includegraphics[scale=0.35,trim=0 0 0 0,clip]{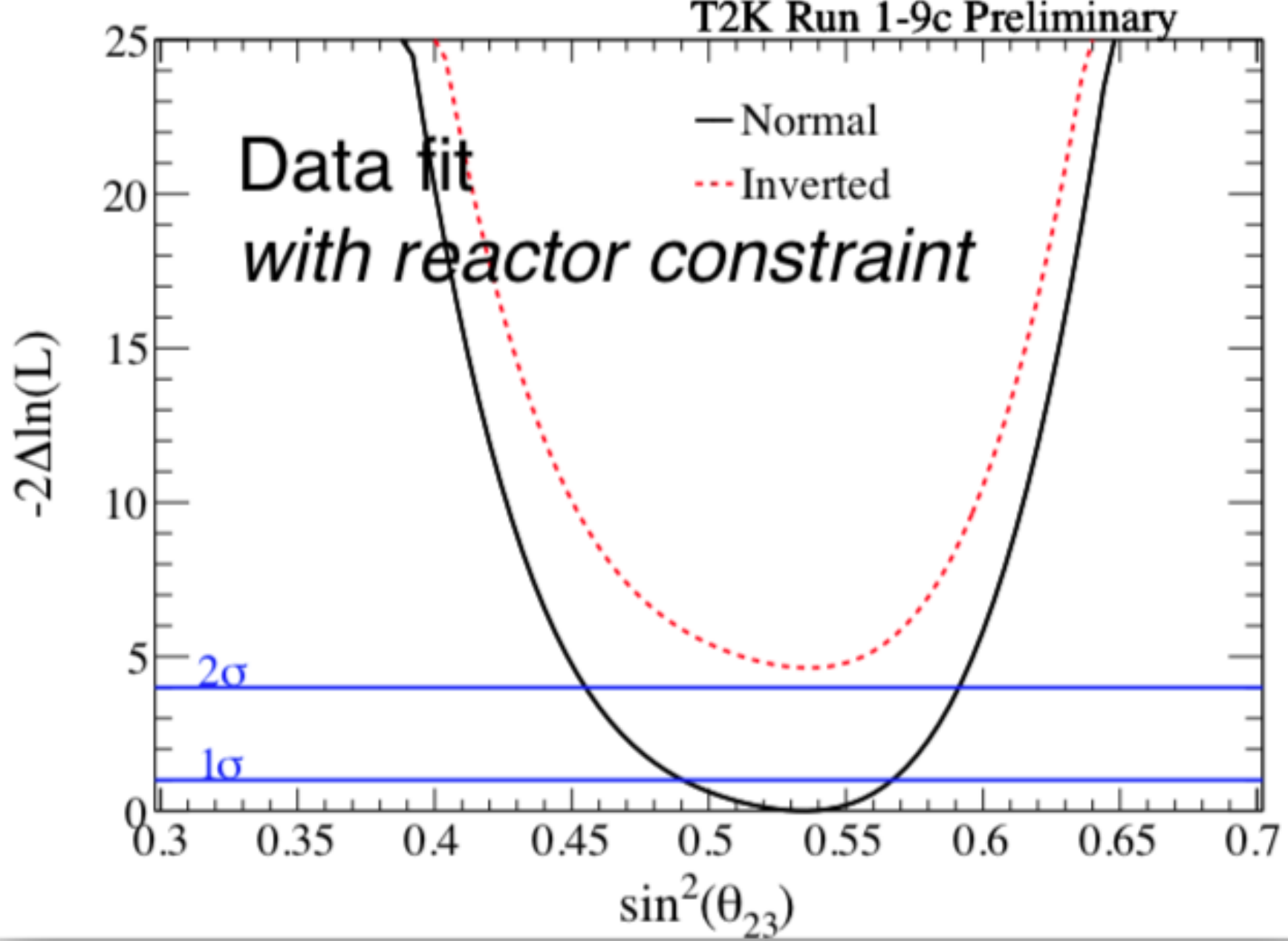}
  \hspace*{7.3cm}\includegraphics[scale=0.35,trim=0 0 0 13,clip]{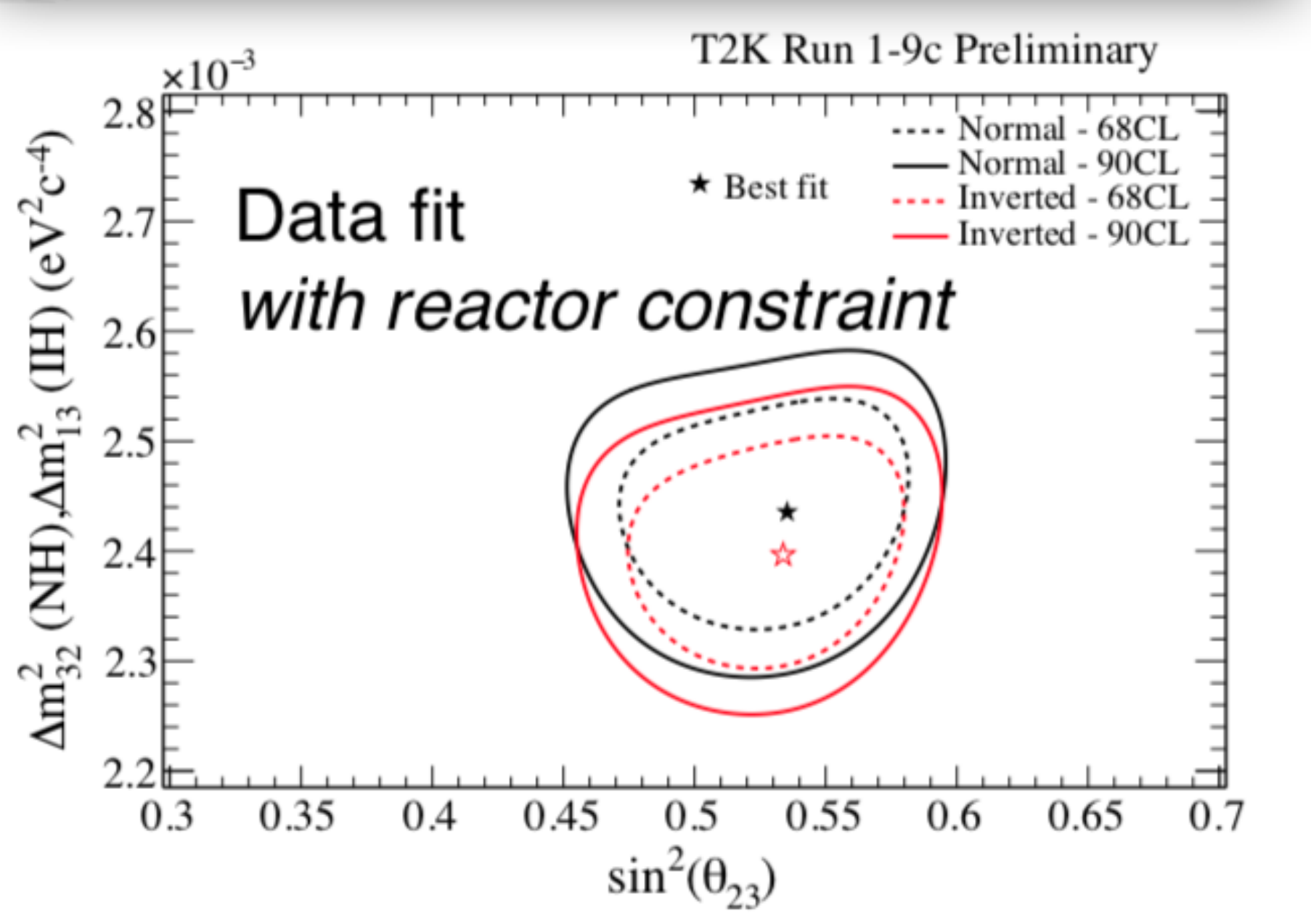}
  \vspace{5.2cm}
  \caption{\label{fig_T2K_contours}
    \emph{Left}. Regions of $\sin^2\theta_{23}$ vs.~$\delta_{CP}$ parameter space obtained by T2K data only (top panel) and combined with reactors (bottom panel). \emph{Right}. Regions of $\Delta m^2_{32}$ vs.~$\sin^2\theta_{23}$ parameter space (top) and confidence intervals for $\sin^2\theta_{23}$ (bottom) obtained by T2K combined with reactor constraint \cite{Wascko:2018Neutrino}.}
\end{center}
\end{figure}

Their data fit analysis was done considering both channels to find constraints on $\Delta m_{23}^2$, $\sin^2\theta_{23}$ and $\delta_{CP}$, but T2K also considered the constraints coming from reactor experiments, and their results are shown in Fig.~\ref{fig_T2K_contours}. The best fit values for the oscillation parameters are
\begin{align}
\Delta |m^2_{32}| &= 2.434\pm0.064 \times 10^{-3} \, \rm{ eV}^2, 
\qquad
& \sin^2\theta_{23} &= 0.536^{+0.031}_{-0.046},
\qquad
& \rm{Normal \, Ordering}; \\
\Delta |m^2_{32}| &= 2.410^{+0.062}_{-0.063} \times 10^{-3} \, \rm{ eV}^2, 
\qquad
&\sin^2\theta_{23} &= 0.536^{+0.031}_{-0.041},
\qquad
& \rm{Inverted \, Ordering}.
\end{align}
Notice that CP conserving values are outside of 2$\sigma$ region for both mass orderings and that normal ordering is favored by data.
\subsection{MINOs and MINOS+}
With its 735 km baseline, MINOS (originally planned to perform research on atmospheric neutrinos in the FD) and MINOS+ were designed to observe muon (anti)neutrino flavor changing from a beam produced at Fermilab and directed towards a two-detector system (the Far detector located at the Soudan Underground Laboratory in Minnesota) \cite{Evans:2017brt}. Thanks to improvements implemented on the NuMI beam, the neutrinos energy peak increased from 3 GeV for MINOS to 7 GeV for MINOS+ \cite{Aurisano:2018Neutrino}.

\begin{figure}[ht]
\begin{center}
\includegraphics [scale=0.18]{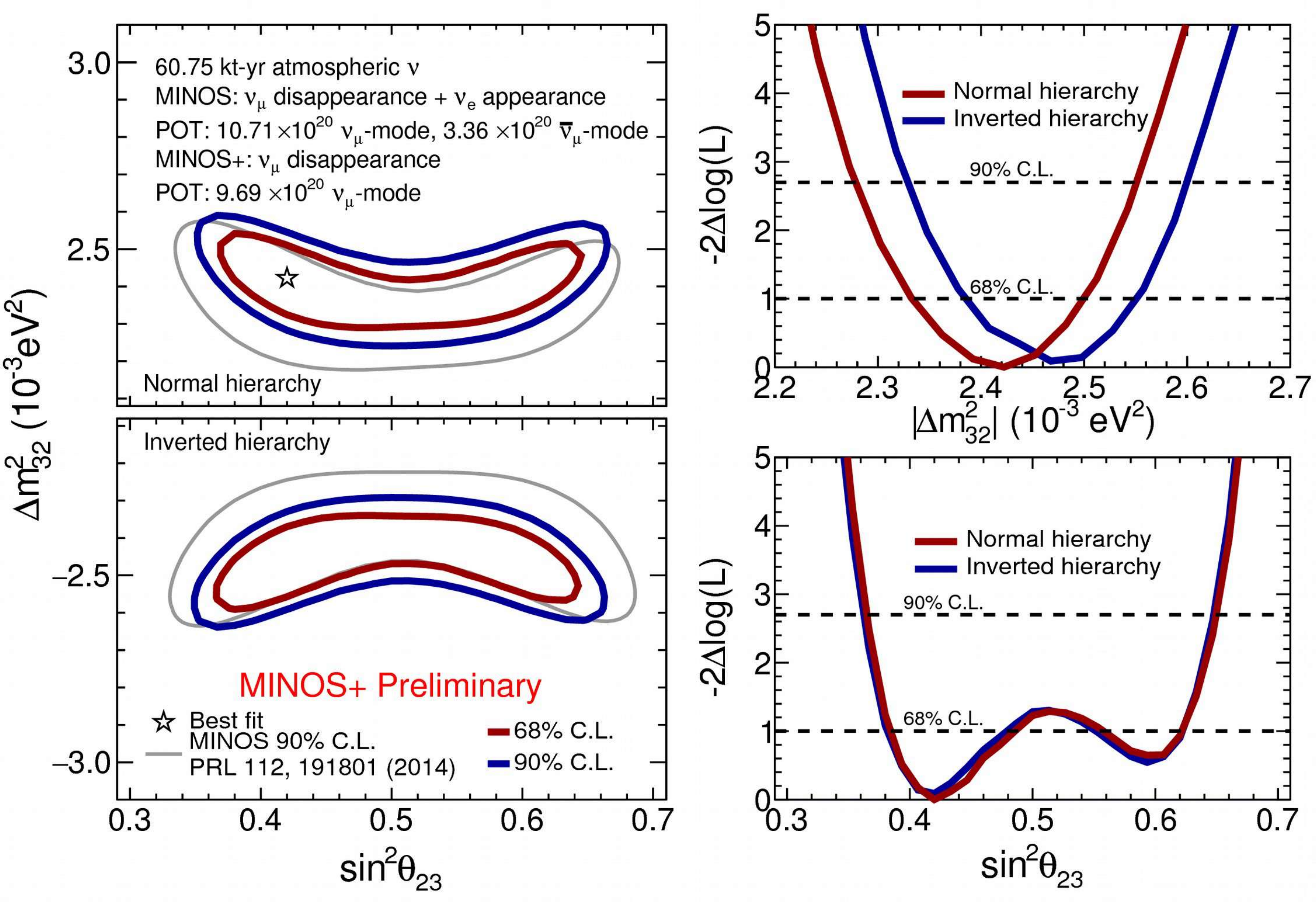}
\caption{\label{fig_MINOS_contours} Regions of $\sin^2\theta_{23}$ vs.~$\delta_{CP}$ parameter space (left) and one-dimensional significance plots for each oscillation parameter (right), obtained by MINOS and MINOS+ data for the Normal and inverted mass orderings \cite{Aurisano:2018Neutrino}.}
\end{center}
\end{figure}

Using neutrinos from the NuMI beam, MINOS+ found that FD data are consistent with three flavor prediction, imposing tightly constrains on alternate oscillations hypotheses \cite{DeRijck:2017ynh,Aurisano:2018Neutrino}. The combination of atmospheric and beam neutrinos (in the appearance and disappearance channels), using data from neutrinos and antineutrinos, allows the collaboration to constraint the oscillation parameters as shown in Fig.~\ref{fig_MINOS_contours}, finding the best fit at 
\begin{equation}
\Delta m^2_{32} = 2.42\times 10^{-3}\,\,\rm{eV}^2,\quad\quad \sin^2\theta_{23} = 0.42. 
\end{equation}
Their results present a 1.1$\sigma$ exclusion of maximal mixing and a 0.8$\sigma$ preference for the lower octant. They also point to a normal mass order preference with a significance of 0.2$\sigma$ \cite{Aurisano:2018Neutrino}.
\subsection{OPERA}
\begin{figure}[!ht]
\begin{center}
\includegraphics [scale=0.5]{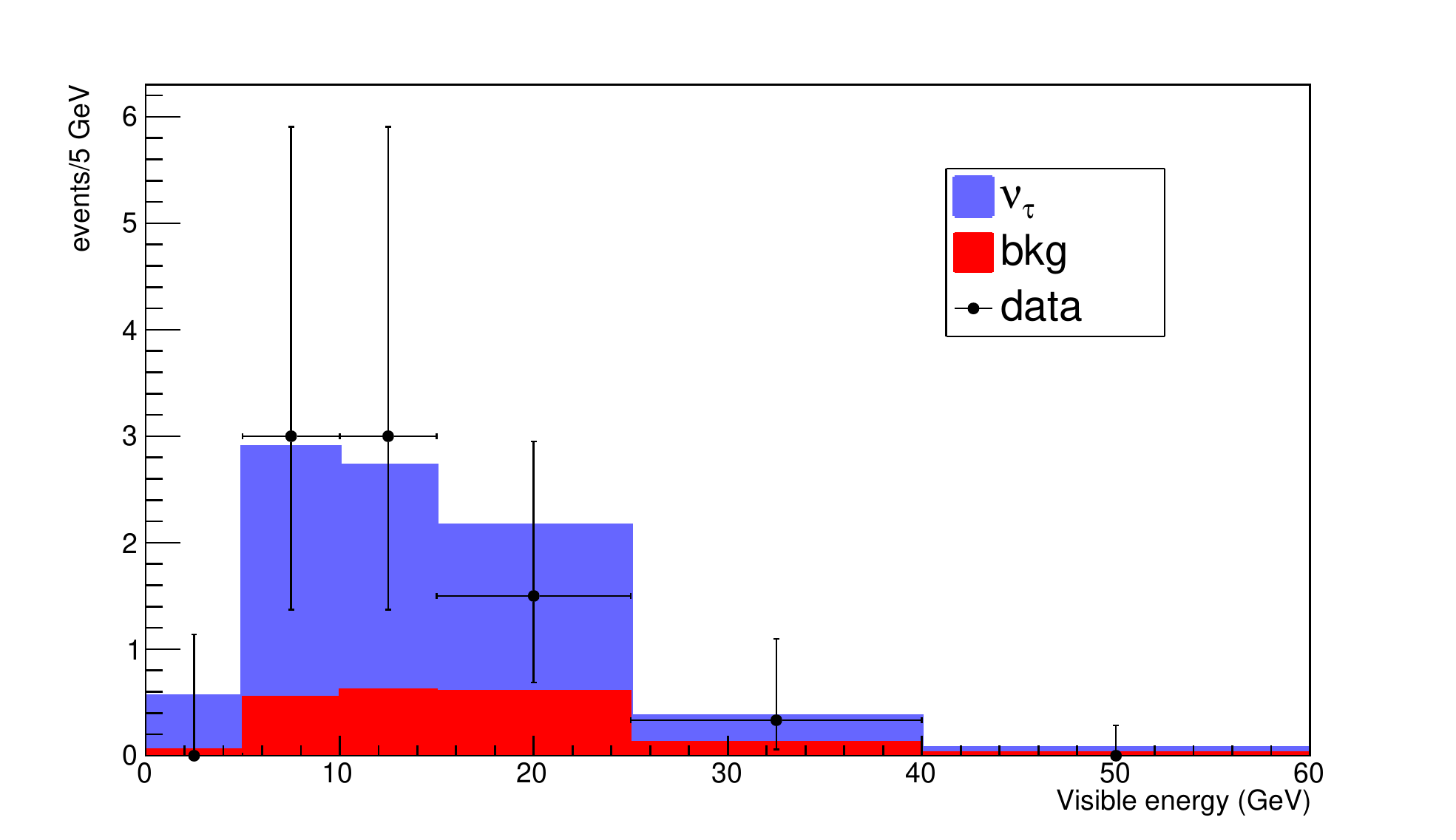}
\caption{\label{fig_OPERA_data} Energy distribution of OPERA data  compared with the expectation \cite{Agafonova:2018auq}.}
\end{center}
\end{figure}
Using a muon-neutrino beam generated at CERN and directed towards a detector located 730 km away, at the LNGS\footnote{Gran Sasso National Laboratory, in Italy}, OPERA \cite{Guler:2000bd} was designed to detect the appearance of tau-neutrinos through the $\nu_{\mu} \to \nu_{\tau}$ oscillation. With a total of 10 $\nu_{\tau}$ candidate events, their final analysis of the full data sample confirms the appearance on tau-neutrinos with a significance of 6.1$\sigma$ (Fig~\ref{fig_OPERA_data}), and the statistical analysis allowed them to report the first measurement of $\Delta m^2_{32}$ from the $\nu_{\tau}$ appearance mode \cite{Agafonova:2018auq}:
\begin{equation}
\Delta m^2_{32} = 2.7^{+0.7}_{-0.6} \times 10^{-3} \,\, {\rm eV}^2,
\end{equation}
which is consistent with the results by other experiments in the disappearance mode.

\section{The Future}
In addition to the expected results coming from the currently active experiments, neutrino oscillations will be extensively end deeply studied by two impressive LBL experiments: Hyper-Kamiokande and DUNE.
\subsection{Hyper-Kamiokande}
\begin{figure}[ht]
\begin{center}
\includegraphics [scale=0.73]{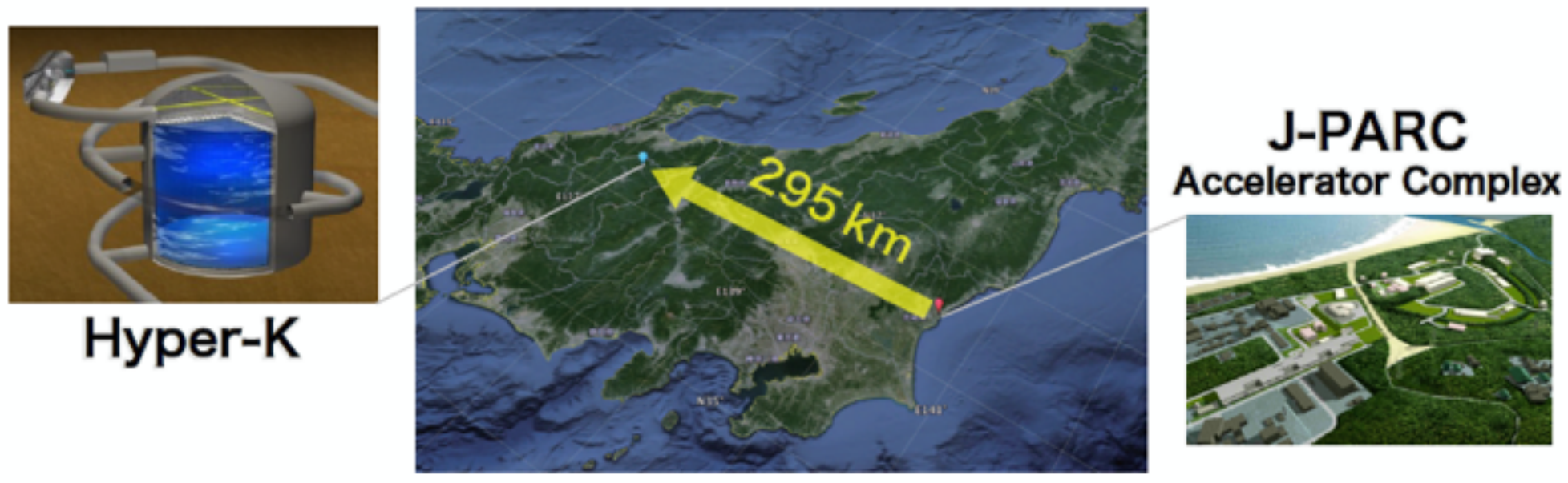}
\caption{\label{fig_HK_exp} The Hyper-Kamiokande experimental layout \cite{Shiozawa:2018Neutrino}.}
\end{center}
\end{figure}
Hyper-Kamiokande (HK), located in Japan, is the successor of and take advantage of all the technological success from the very well known Super-Kamiokande (SK) experiment. This water Cherenkov detector will be placed in the Tochibora mine, about 295 km away from the J-PARC proton accelerator research complex in Tokai \cite{Abe:2018uyc} (Figure \ref{fig_HK_exp}).
\begin{figure}[ht]
\begin{center}
\includegraphics [scale=0.37]{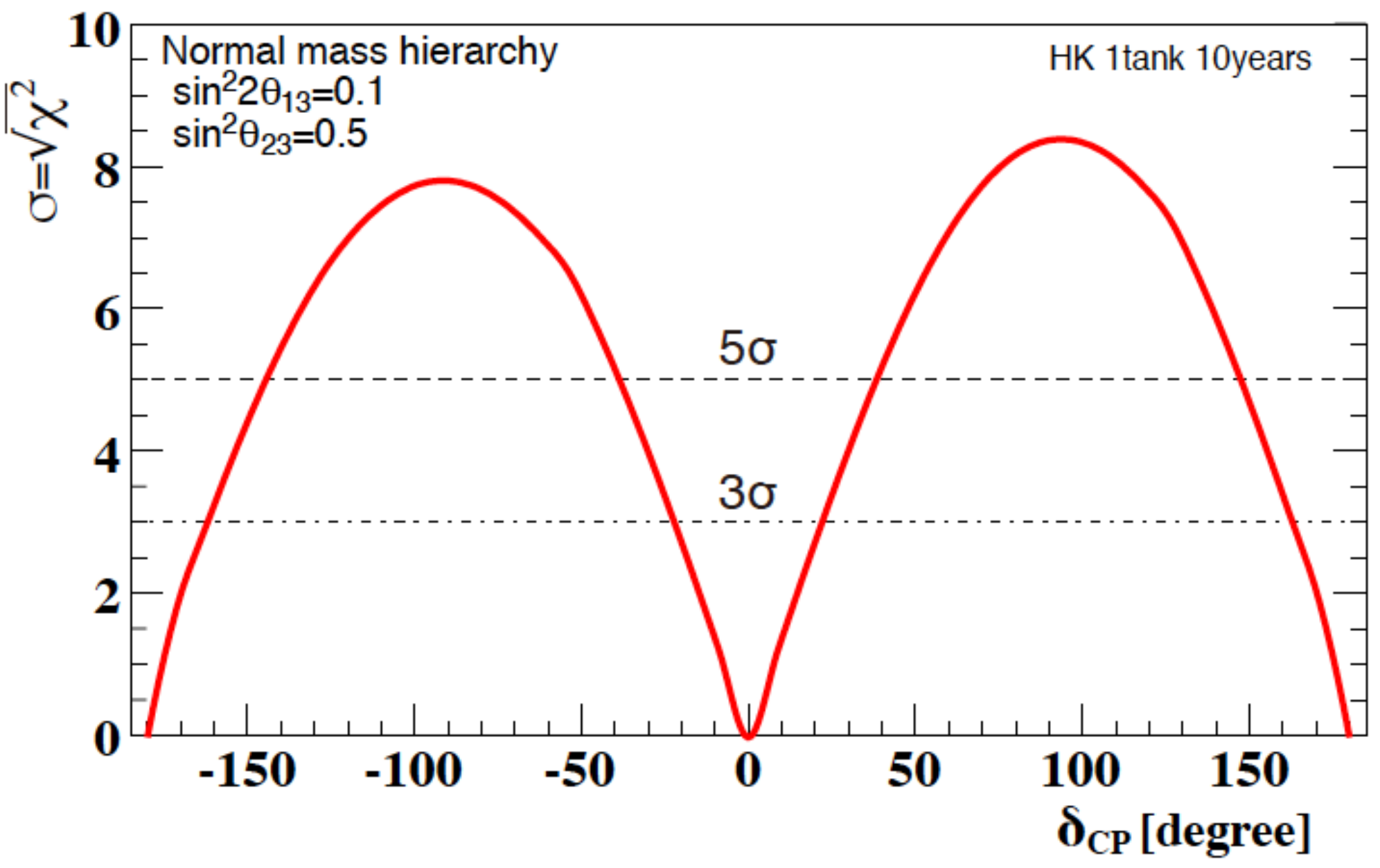}
\includegraphics [scale=0.37]{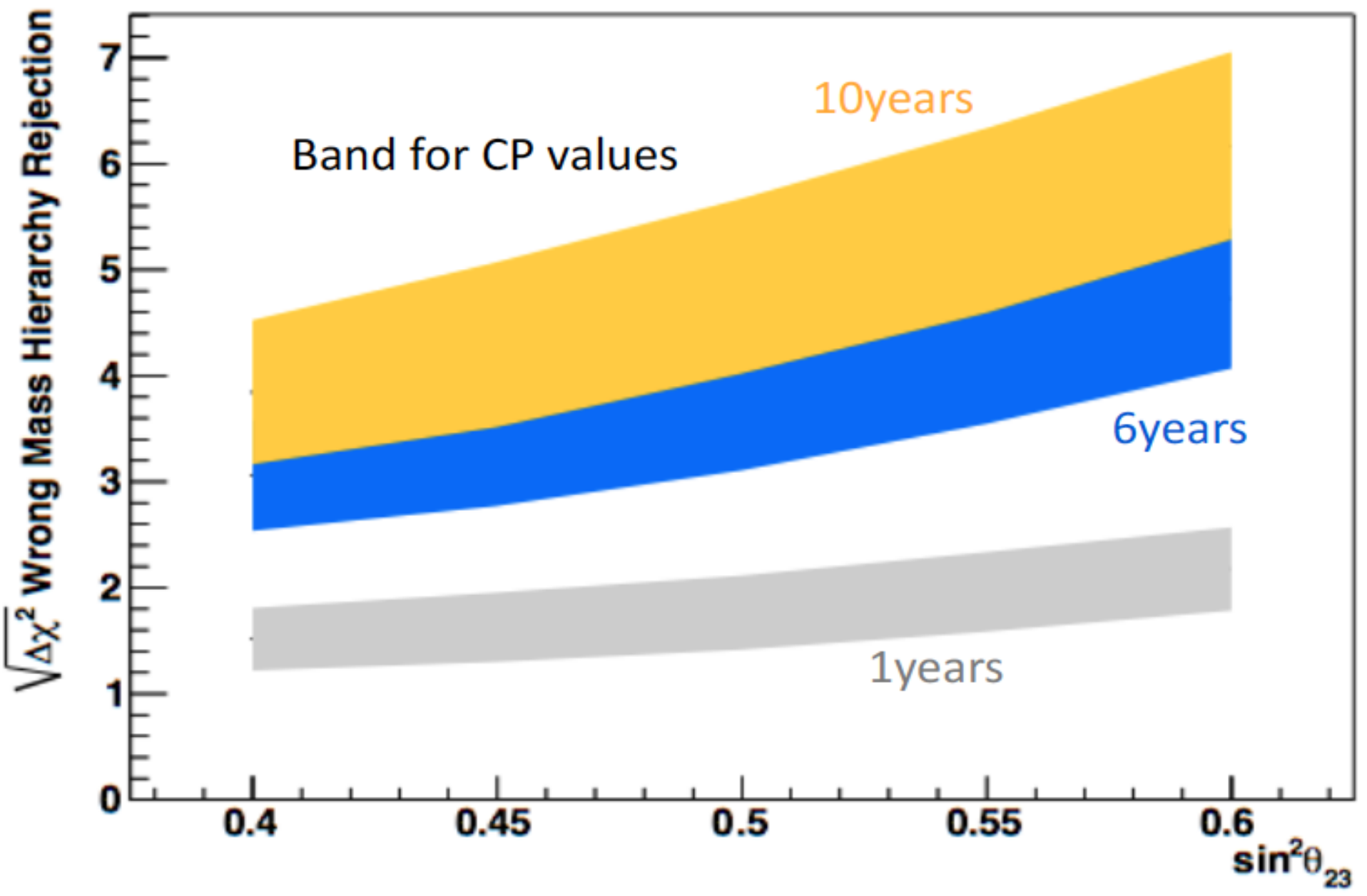}
\caption{\label{fig_HK_cpv} Expected significance to exclude $\delta_{CP} = 0$ for the normal mass order (\emph{Left panel}) and neutrino Mass Order sensitivity as a function of the true value of $\sin^2\theta_{23}$ (\emph{Right panel}) in the HK experiment \cite{Abe:2018uyc,Shiozawa:2018Neutrino}.}
\end{center}
\end{figure}
As can be seen in Fig.~\ref{fig_HK_cpv} (left panel), HK has the potential to exclude CP conservation ($\delta_{CP} = 0$) for the normal mass ordering, with a significance larger than $5\sigma$ after 10 years of data-taking. With enough time, HK will also reach a large sensitivity for the determination of the correct neutrino mass ordering (depending on the value of $\delta_{CP}$ (right panel of Fig.~\ref{fig_HK_cpv}).

However, HK has a rich scientific program which goes beyond the study od neutrino oscillations, including the search of nucleon decays, neutrinos emitted by supernova and from other astrophysical sources (dark matter annihilation, gamma ray burst jets, and pulsar winds) \cite{Abe:2018uyc}.
\subsection{DUNE}
\begin{figure}[ht]
\begin{center}
\includegraphics [scale=0.55]{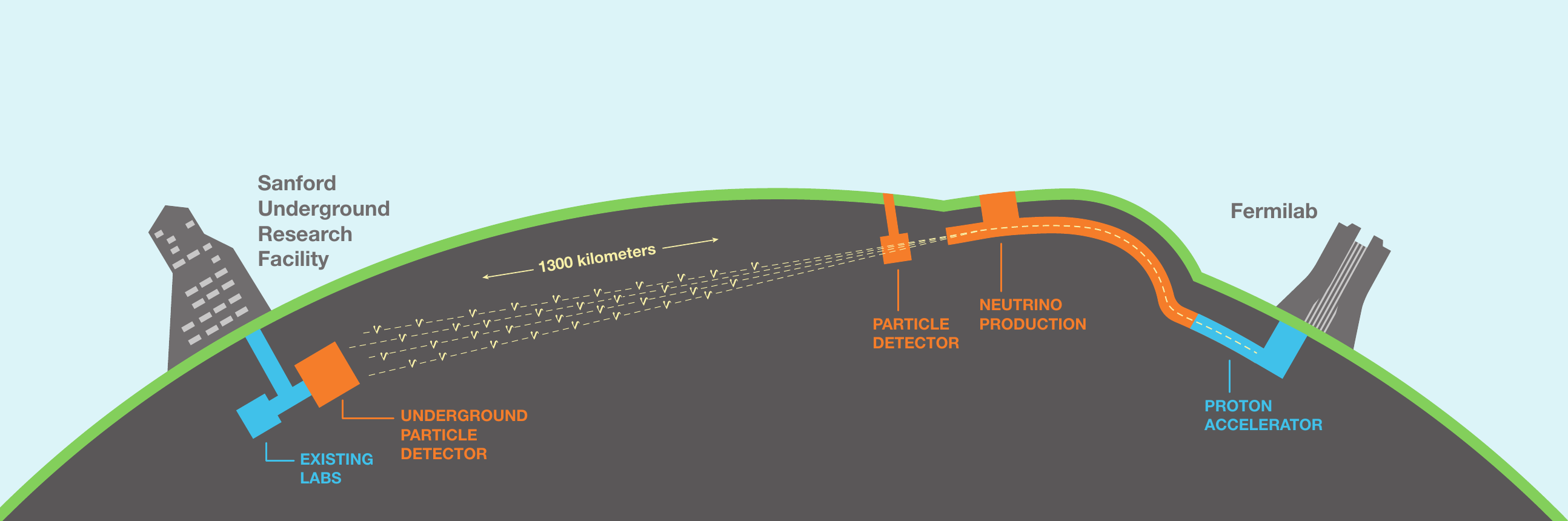}
\caption{\label{fig_DUNE_exp} The DUNE experimental layout.}
\end{center}
\end{figure}
The Deep Underground Neutrino Experiment (Figure \ref{fig_DUNE_exp}) \cite{Acciarri:2016crz} will be based at Fermilab, from where an intense neutrino beam is going to be fired towards a system of two Liquid Argon detectors, 1300 km apart: the Near Detector (ND), placed at Fermilab and the Far Detector (FD) at the Sanford Underground Research Facility.
\begin{figure}[!ht]
\begin{center}
\includegraphics [scale=0.37]{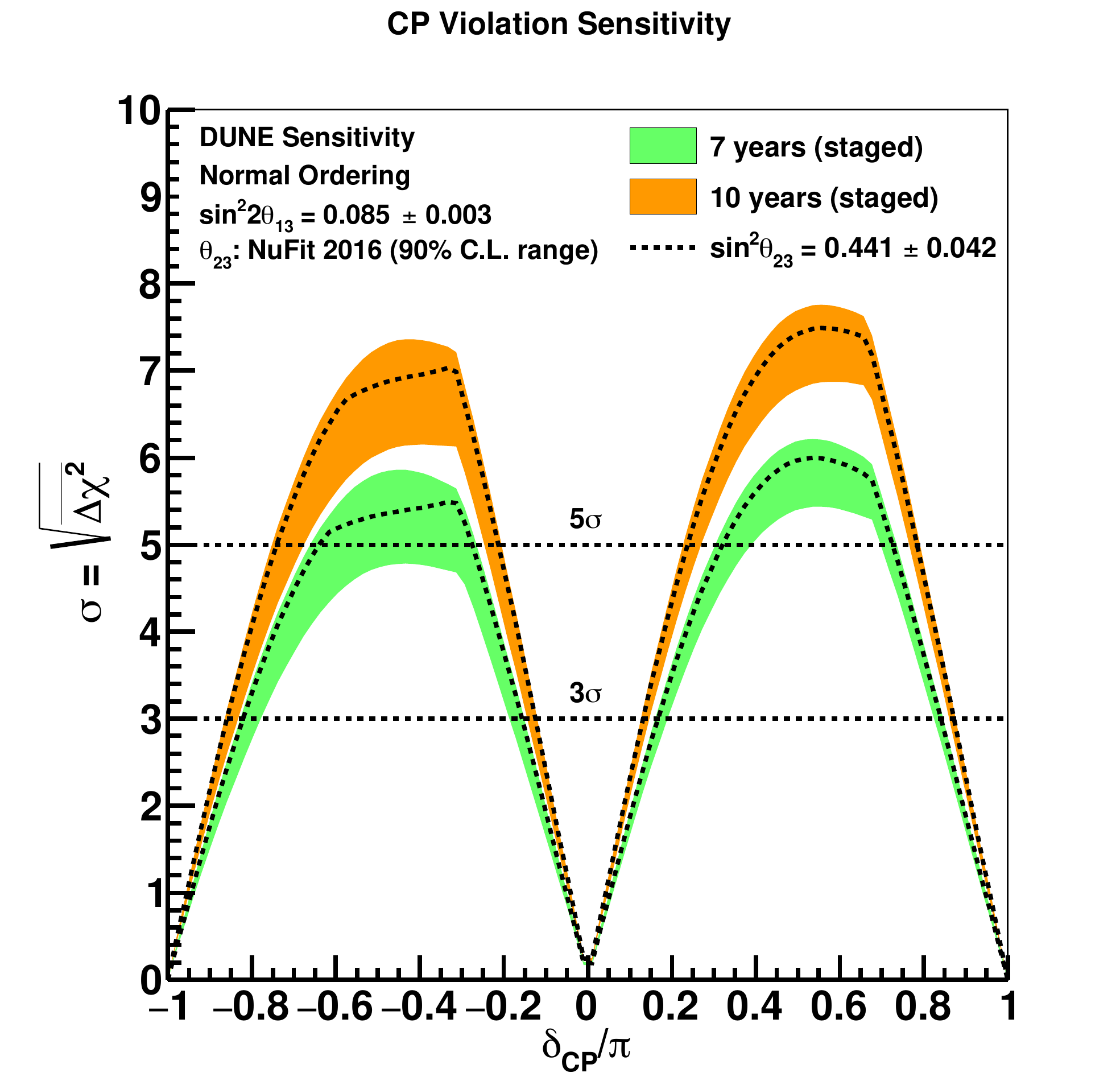}
\includegraphics [scale=0.37]{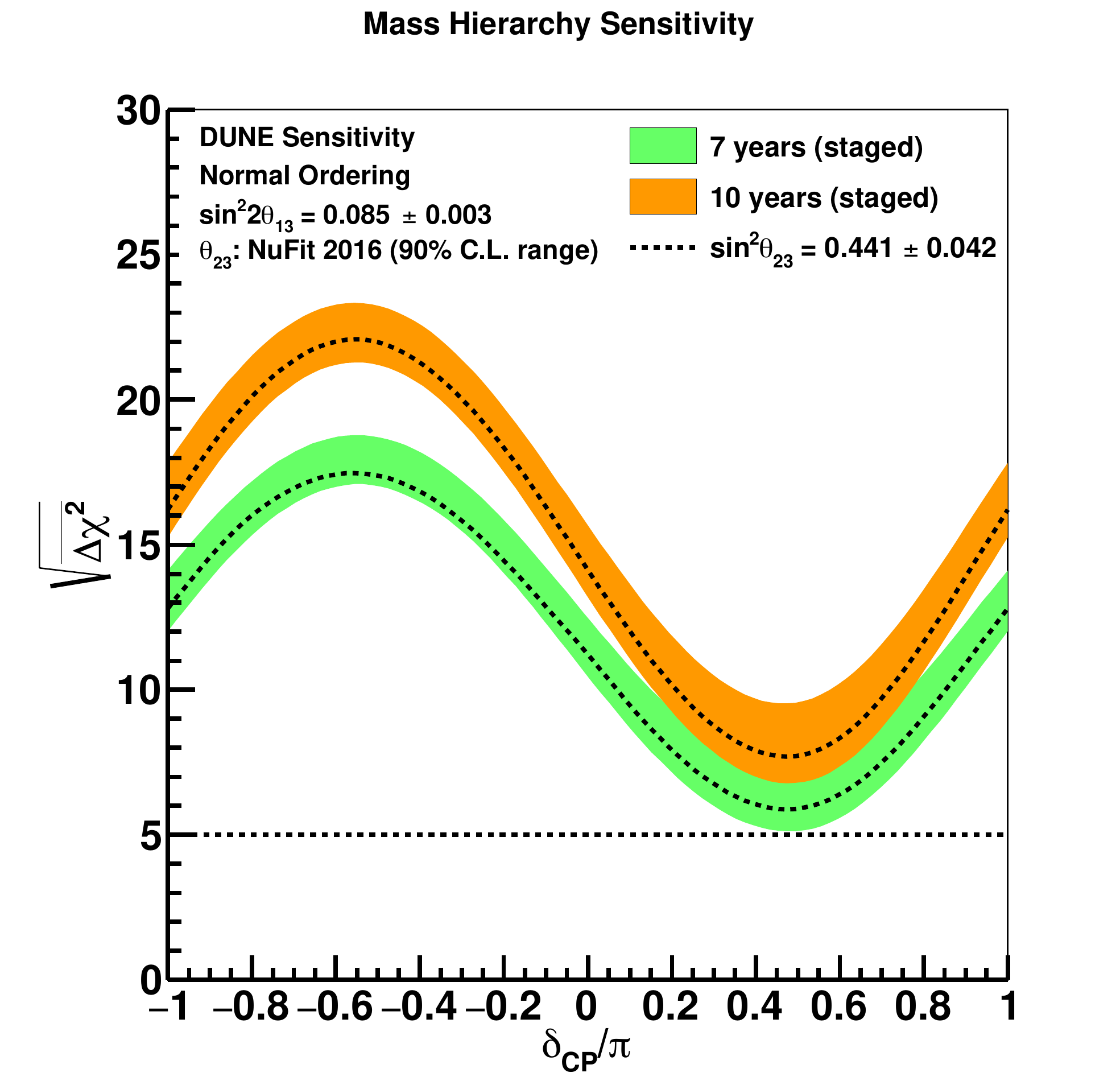}
\caption{\label{fig_DUNE_cpv} Significance with which CP violation (\emph{Left panel}) and the neutrino Mass Order (\emph{Right panel}) can be determined as a function of the true value of $\delta_{CP}$, for exposures of seven (green) or 10 (orange) years, and for the Normal Mass Ordering case in the DUNE experiment. The width of the band represents the range of sensitivities for the 90\% C.L. range of $\theta_{23}$ values \cite{Brailsford:2018dzn}.}
\end{center}
\end{figure}

DUNE is expected to explore the neutrino oscillation phenomenon with an unprecedented precision, aiming to determine the mass ordering and the value of the CP-violating phase, $\delta_{CP}$. As depicted in Fig.~\ref{fig_DUNE_cpv}, DUNE should be capable of determining the CP violation (i.e. measuring $\delta_{CP} \neq 0,\pi$)with a significance around $5\sigma$ after 7 years of data-taking, and even larger significance after 10 years. For the neutrino Mass Ordering (right panel of Fig.~\ref{fig_DUNE_cpv}), after 7 years, DUNE could reach a $5\sigma$ sensitivity for all possibe values of $\delta_{CP}$ \cite{Brailsford:2018dzn,Worcester:2018Neutrino}.

In addition, DUNE will be able to search for signals of proton decay, neutrinos coming form supernovae and some exotic physics related to sterile neutrinos, non-standard neutrino interactions and Dark Matter, for instance \cite{Acciarri:2015uup}.
\section{Conclusions}
Research on neutrino physics, and specially on neutrino oscillations, has been extensive and we are currently living a very exciting time with very important observations and with increasing precision measurements: strong evidence of $\bar{\nu}_e$ and $\nu_{\tau}$ appearance, CP-conserving values excluded at 2$\sigma$, data preference of normal mass ordering. All these confirming the 3 flavor oscillation hypothesis.

On the other hand, as there are yet opened questions to be resolved (precise measurement of $\delta_{CP}$; existence of sterile neutrinos; the Dirac/Majorana nature and the absolute mass of neutrinos, among others), a number of proposed experiments are starting to become real as they are at their commissioning and/or building stages, allowing us to foresee a bright future for the neutrino physics scientific community and beyond.

\section*{Acknowledgment}

I am deeply grateful to the Organizers of the PIC2018 symposium for the kind invitation to participate in this event, and to the NOvA Collaboration for their scientific support. I also thank the \emph{Vicerrector\'ia de Investigaciones, Extensi\'on y Proyecci\'on Social} of the Universidad del Atl\'antico for their financial support.


\end{document}